\documentclass[amsmath,amssymb,prr,reprint,superscriptaddress,nofootinbib]{revtex4-2}
\usepackage {graphicx}  
\usepackage{dcolumn}  
\usepackage{bm}  
\usepackage[whole]{bxcjkjatype}   
\usepackage{verbatim}  
\usepackage{color} 
\usepackage{xcolor}  
\usepackage{physics}   
\usepackage{caption}  
\usepackage{subcaption}
\usepackage[position=top,singlelinecheck=false]{subcaption}

\begin{document}
\title{Learning parameter dependence for Fourier-based option pricing with tensor trains}

\author{Rihito Sakurai}
\email{sakurairihito@gmail.com}
\affiliation{Department of Physics, The University of Tokyo, Tokyo 113-0033. Japan}
\affiliation{Department of Physics, Saitama University, Saitama 338-8570, Japan}

\author{Haruto Takahashi}
\affiliation{Department of Physics, Saitama University, Saitama 338-8570, Japan}

\author{Koichi Miyamoto}
\affiliation{Center for Quantum Information and Quantum Biology, Osaka University, Osaka 560-0043, Japan}

\date{\today}

\begin{abstract}
A long-standing issue in mathematical finance is the speed-up of option pricing, especially for multi-asset options. 
A recent study has proposed to use tensor train learning algorithms to speed up Fourier transform (FT)-based option pricing, utilizing the ability of tensor trains to compress high-dimensional tensors. 
Another usage of the tensor train is to compress functions, including their parameter dependence. 
Here, we propose a pricing method, where, by a tensor train learning algorithm, we build tensor trains that approximate functions appearing in FT-based option pricing with their parameter dependence and efficiently calculate the option price for the varying input parameters.
As a benchmark test, we run the proposed method to price a multi-asset option for the various values of volatilities and present asset prices.
We show that, in the tested cases involving up to 11 assets, the proposed method outperforms Monte Carlo-based option pricing with $10^5$ and $10^6$ paths in terms of computational complexity while keeping comparable accuracy.
\end{abstract}

\maketitle

\section{Introduction}
Financial firms are conducting demanding numerical calculations in their business.
One of the most prominent ones is option pricing.
An option is a financial contract in which one party, upon specific conditions being met, pays an amount (payoff) determined by the prices of underlying assets such as stocks and bonds to the other party.
Concerning the time when the payoff occurs, the simplest and most common type of options is the European type, which this paper hereafter focuses on: at the predetermined future time (maturity) $T$, the payoff $v(\vec{S}(T))$ depending on the underlying asset prices $\vec{S}(T)$ at time $T$ occurs.

For example, in a single-asset European call/put option, one party has the right to buy/sell an asset at the predetermined price (strike) $K$ and maturity $T$, and the corresponding payoff function is $v(S(T))=\max\{c(S(T)-K),0\}$, where $c=1$ and $-1$ for a call and put option, respectively.
In addition to this simple one, various types of options are traded and constitute a large part of the financial industry.

Pricing options appropriately is needed for making a profit and managing the risk of loss in option trading.
According to the theory in mathematical finance \footnote{As typical textbooks in this area, we refer to Refs.~\cite{hull2003options,shreve2004stochastic}.}, the price of an option is given by the expectation of the discounted payoff in the contract with some stochastic model on the dynamics of underlying asset prices assumed.
Except for limited cases with simple contract conditions and models, the analytical formula for the option price is not available and thus we need to resort to the numerical calculation.
In the rapidly changing financial market, quick and accurate pricing is vital in option trading, but it is a challenging task, for which long-lasting research has been made.
In particular, pricing multi-asset options, whose payoff depends on the prices of multiple underlying assets, is often demanding.
Pricing methodologies such as PDE-based solvers (e.g., the finite difference method~\cite{duffy2013finite}) and 
tree models~\cite{Clewlow1998} suffer from the so-called curse of dimensionality, which means the exponential increase of computational complexity with respect to the asset number, and the Monte Carlo method, which may evade the exponential complexity, has a slow convergence rate.
The quasi-Monte Carlo method is more effective than the Monte Carlo method up to a certain level of the dimension but has the worst-case complexity exponential with respect to the dimension, despite the methods for mitigating this proposed~\cite{Bayer04032018,Liu2023,bayer2024quasimontecarloefficientfourier}.
At any rate, in the rapidly changing financial market, the pursuit of the speed-up of option pricing can never go too far.

Motivated by these points, recently, applications of quantum computing to option pricing are considered actively\footnote{See Ref.~\cite{herman2023quantum} as a comprehensive review.}.
For example, many studies have focused on applications of the quantum algorithm for Monte Carlo integration \cite{montanaro2015}, which provides the quadratic quantum speed-up over the classical counterpart.
Unfortunately, running such a quantum algorithm requires fault-tolerant quantum computers, which may take decades to be developed. 

In light of this, applications of quantum-inspired classical algorithms, that is tensor network (TN) algorithms, to option pricing have also been studied as solutions in the present ~\cite{doi:10.1137/19M1244172,kastoryano2022highlyefficienttensornetwork,patel2022quantuminspired, doi:10.1137/21M1402170}.
Among them, this paper focuses on the application of tensor train (TT)~\cite{Oseledets2011}learning to the FT-based option pricing method, following the original proposal in Ref.~\cite{kastoryano2022highlyefficienttensornetwork}.
This option pricing method is based on converting the integration for the expected payoff in the space of the asset prices $\vec{S}(T)$ to that in the Fourier space, namely, the space of $\vec{z}$, the wavenumbers corresponding to the logarithm of $\vec{S}(T)$. 
After this conversion, the numerical integration is done more efficiently in many cases.
Unfortunately, the FT-based method with naive grid-based integration suffers from the curse of dimensionality, with its computational complexity increasing exponentially for multi-asset options.
On the other hand, tensor network is the technique originally developed in quantum many-body physics to express state vectors with exponentially large dimension efficiently\footnote{See \cite{ORUS2014117,Okunishi2022} as reviews.} and recently, it has also been utilized in various fields such as machine learning \cite{Stoudenmire2016,novikov2016exponential,sozykin2022ttopt}, quantum field theory~\cite{Shinaoka2023-lf, Nunez_Fernandez2022-fo,takahashi2024compactness}, and partial differential equations~\cite{patel2022quantuminspired, PhysRevE.106.035208, kornev2023numerical, Gourianov_2022}.
Ref.~\cite{kastoryano2022highlyefficienttensornetwork} proposed to leverage the ability of tensor network to compress data as a high-dimensional tensor in order to express the functions of $\vec{z}$ involved in FT-based option pricing~\footnote{As another approach for accelerating FT-based multi-asset option pricing, see~\cite{Bayer_2023}.}.
The authors built TTs, a kind of tensor network, approximating those functions by a TT learning algorithm called tensor cross interpolation (TCI)~\cite{Oseledets2010-fg,Dolgov2020-yi, Nunez_Fernandez2022-fo, Ritter2024-pr} and evaluated the integral involving them efficiently, which led to the significant speed-up of FT-based option pricing in their test cases.

Here, we would like to point out an issue in this TT-based method.
That is, we need to rerun TCI to obtain TTs and compute option price each time the input parameters such as volatility and initial asset price, are changed.
According to our numerical experiments, the TT learning method for FT-based option pricing~\cite{kastoryano2022highlyefficienttensornetwork} takes a longer time than the Monte Carlo method, which means that the TT-based method does not have a computational time advantage.

To address this issue, we focus on another usage of tensor trains that can embed parameter dependence~\cite{990454, 10.1007/3-540-47969-4_30,10.1063/5.0045521,Ballani2015} to make FT-based option pricing more efficient.
Namely, we learn TTs that approximate the functions including not only the dependence on  $\vec{z}$ but also that on parameters in the asset price model such as the volatilities and the present asset prices, by a single application of TCI for each function.
We use these tensor trains to evaluate the integral including parameter dependence and perform fast option pricing in response to various parameter changes (refer to FIG.~\ref{fig:new_scheme}).
Note that this is an advantage of the proposed method not only over the original TT-bsed method in Ref.~\cite{kastoryano2022highlyefficienttensornetwork} but also over other aforementioned methods such as PDE-based solvers, tree models, and the (quasi-)Monte Carlo method.
When parameters are changed and an option is repriced, these methods must be rerun from scratch, in contrast to the proposed method, in which we just reuse the TT that is evaluated efficiently.

For evaluation, we consider two benchmark scenarios in which we vary volatilities and present stock prices under the Black–Scholes model, focusing specifically on a min-option. 
In the test cases, it is seen that for up to 11 assets, the computational complexity of our proposed method, which is measured by the number of elementary operations is advantageous to that of the Monte Carlo method with $10^6$ paths by a factor of  $O(10^5)$ (see Fig.~\ref{fig:table_cc}).
We also confirm numerically that in the tested cases, the accuracy of our method is within the statistical error in the Monte Carlo method with $10^6$ paths.
In summary, these results indicate that, at least in the tested cases, 
our proposed method offers significant advantages in computational complexity while keeping a better accuracy.

In the context of TT-based approximations of high-dimensional functions, incorporating parameter dependence into TTs has been considered in some fields~\cite{990454, 10.1007/3-540-47969-4_30,10.1063/5.0045521,Ballani2015}.
However, to the best of our knowledge, our study is the first to take such an approach in FT-based option pricing in order to make it more efficient for varying parameters, which provides practical benefits in the rapidly changing financial market.

\section{Tensor train}
A $d$-way tensor $F_{x_1, \dots, x_d}$, where each local index $x_l$, $l=1, \dots, d$, has a local dimension $N$, can be decomposed into a TT format with a low-rank structure. 
The TT decomposition of $F_{x_1, \ldots, x_d}$ can be expressed as follows. 
\begin{align}
    F_{x_1, \ldots, x_d} &\approx \sum_{l_1}^{\chi_1} \cdots \sum_{l_{d-1}}^{\chi_d} F_{l_1, x_1}^{(1)} F_{l_1 l_2, x_2}^{(2)} \cdots F_{l_{d-1}, x_d}^{(d)} \notag \\
    &\equiv 
    \prod_{i=1}^{d} F_{x_i}^{(i)}
\end{align}
where $F_{x_i}^{(i)}$ denotes each 3-way tensor, $l_i$ represents the virtual bond index, and $\chi_i$ is the dimension of the virtual bond. 
One of the main advantages of TT is that it significantly reduces computational complexity and memory requirements by reducing bond dimensions $\chi_i$. 
Tensor train is a mathematically equivalent to matrix product state (MPS)~\cite{Schollwock2011-eq}.

This is an equivalent expression to the wave function $F_{x_1, \ldots, x_d}$ of a quantum system with $d$ $N$-level qudits as follows:
\begin{align}
    \ket{F_{x_1, x_2,\ldots, x_d}} = \sum_{l_1}^{\chi_1} \cdots \sum_{l_{d-1}}^{\chi_d} F_{l_1, x_1}^{(1)} F_{l_1 l_2, x_2}^{(2)} \cdots F_{l_{d-1}, x_d}^{(d)} 
    \ket{\vec{x}},
\end{align}
where $\ket{\vec{x}}=\ket{x_1}\cdots\ket{x_d}$ is the tensor product of $\ket{x_1},\cdots,\ket{x_d}$, the basis states from $\ket{1}$ to $\ket{N}$.

In the same manner, the $2d$-way tensor $F_{x_1, \dots, x_d}^{y_1, \dots, y_d}$ where each local index $x_l$ has local dimension $N_l$ and each local index $y_l$ has local dimension $M_l$, $l=1, \dots, d$, can be expressed with the product of the each tensor core, i.e., fourth-order tensor $[F_{a_{i-1}x_i a_{i}}^{y_i}]^{(i)}$,
\begin{align}
    F_{x_1, \ldots, x_d}^{y_1, \ldots, y_d} 
    &\approx 
    \sum_{a_1=1}^{\chi_1} \cdots \sum_{a_{d-1}=1}^{\chi_d} 
    [F_{a_0 x_1 a_1}^{y_1}]^{(1)}  \cdots [F_{a_{d-1} x_d a_d}^{y_d}]^{(d)}  \nonumber \\
    &= \prod_{i=1}^{d}
    [F_{x_i }^{y_i}]^{(i)}
\end{align}
is called the tensor train operator (TTO) or matrix product operator (MPO)~\cite{Schollwock2011-eq}.

\subsection{Compression tenchniques}\label{subsec:comp}
We introduce the two compression techniques used in this study.

\subsubsection{Tensor cross interpolation}
Tensor cross interpolation (TCI) is a technique to compress tensors corresponding to
discretized multivariate functions with a low-rank TT representation.
Here, we consider a tensor that, with grid points set in $\mathbb{R}^d$, has entries $F_{x_1,\ldots,x_d}$ equal to $F(x_1,\ldots,x_d)$, the values of a function $F$ on the grid points\footnote{Although we here denote the indexes of the tensor and the variables of the function by the same symbols $x_1,\ldots,x_d$ for illustrative presentation, we assume that, in reality, the grid points in $\mathbb{R}^d$ is labeled by integers and the indexes of the tensor denotes the integers.}.
Leaving the detailed explanation to Refs. \cite{Oseledets2010-fg,Dolgov2020-yi, Nunez_Fernandez2022-fo, Ritter2024-pr}, we describe its outline.
It learns a TT using the values of the target function $F_{x_1, x_2\ldots, x_d}$ at indexes $(x_1,x_2,\cdots,x_d)$ adaptively sampled according to the specific rules. 
TCI actively inserts adaptively chosen interpolation points (pivots) from the sample points to learn the TT, which can be seen as a type of active learning.
It gives the estimated values of the function at points across the entire domain although we use only the function values at a small number of sample points in learning.
This is the very advantage of TCI and is particularly useful for compressing target tensors with a vast number of elements, contrary to singular value decomposition (SVD) requiring access to the full tensor.
Note that TCI is a heuristic method, which means its effectiveness heavily depends on the internal algorithm to choose the pivots and the initial set of points selected randomly.

In this study, when we learn TT from functions with TCI, we add the pivots so that the error in the maximum norm $(\epsilon_{\mathrm{TCI}})$ is to be minimized.
\begin{align}
    \epsilon_{\mathrm{max}} = \frac{\|F_{x_1, x_2,\ldots, x_d}-\tilde{F}_{\mathrm{TT}}\|_{\mathrm{max}}}{\|F_{x_1, x_2,\ldots, x_d}\|_{\mathrm{max}}}
    \label{eq:errTCI}
\end{align}
where $F_{x_1, x_2,\ldots, x_d}$ is a target tensor\footnote{Here, we assume that we can access the arbitrary elements of the target tensor. Note that we do not need to store all the elements of the target tensor.}, $\tilde F_\mathrm{TT}$ is a low-rank approximation, and the maximum norm is evaluated as the maximum of the absolute values of the entries at the pivots selected already.
The computational complexity of TCI is roughly proportional to the number of elements in the TT, which is $O(d \chi^2 N)$ with $\chi_1,\cdots,\chi_d$ fixed to $\chi$. 
In addition, considering the case that zero is included in the reference function value, the error should be normalized by $\|F_{x_1, x_2,\ldots, x_d}\|_{\mathrm{max}}$.

\subsubsection{Singular value decomposition}
In this study, we use singular value decomposition (SVD) to compress further the TTs obtained by TCI with its error threshold $\epsilon_{\mathrm{TCI}}$ set to a sufficiently low.
This is done by first canonicalizing the TT using QR decompositions, and then performing the compression via SVD, discarding singular values that are smaller than the tolerance $\epsilon_{\mathrm{SVD}}$ set for each bond.
This tolerance is defined by
\begin{align}
\epsilon_{\mathrm{SVD}}
=
\frac{|\tilde{F}_{\mathrm{TT}}-\Tilde{F}^{'}_{\mathrm{TT}}|^2_{\mathrm{F}}}{|\tilde{F}_{\mathrm{TT}}|^2_{\mathrm{F}}},
\label{eq:epsilon_SVD}
\end{align}
where $|\cdots|^2_{\mathrm{F}}$ indicates the Frobenius norm, $\tilde{F}_{\mathrm{TT}}$ is the TT obtained from TCI and $\Tilde{F}^{'}_{\mathrm{TT}}$ is the other TT after SVD.
For more technical details, readers are referred to Ref.~\cite{Schollwock2011-eq, Oseledets2011}.

\section{Fourier transform-based option pricing aided by tensor cross interpolation}

\subsection{Fourier transform-based option pricing}

In this paper, we consider the underlying asset prices $\vec{S}(t)=(S_1(t),\cdots,S_d(t))$ in the Black-Scholes (BS) model described by the following stochastic differential equation
\begin{align}
    dS_{m}(t) = rS_{m}(t)dt + \sigma_m S_{m}(t) dW_{m}(t).
\label{black and scholes multi assets}
\end{align}
Here, $W_1(t),\cdots,W_d(t)$ are the Brownian motions with constant correlation matrix $\rho_{ij}$, namely
\begin{align}
    dW_{m}(t) dW_{n}(t) = \rho_{mn}dt, \tag{10}
\label{gauss correlation}
\end{align}
where $r\in\mathbb{R}$ and $\sigma_1,\cdots,\sigma_d>0$ are constant parameters called the risk-free interest rate and the volatilities, respectively.
The present time is set to $t=0$ and the present asset prices are denoted by $\vec{S}_0=(S_{1,0},\cdots,S_{d,0})$.

We consider European-type options, in which the payoff $v(\vec{S}(T))$ depending on the asset prices $\vec{S}(T)$ at the maturity $T$ occurs at $T$.
According to the theory of option pricing, the price $V$ of such an option is given by the expectation of the discounted payoff:
\begin{align}
    V(\vec{p})&=\mathbb{E}\left[e^{-rT}v(\vec{S}(T))\middle|\vec{S}_0\right] \nonumber \\
    &=e^{-rT} \int_{-\infty}^{\infty} v(\exp(\vec{x}))q(\vec{x}|\vec{x}_0)dx,
    \label{expec val}
\end{align}
where we define $\exp(\vec{x}):=(e^{x_1},\cdots,e^{x_d})$.
$q(\vec{x}|\vec{x}_0)$ is the probability density function of $\vec{x}:=(\log S_1(T),\cdots,\log S_d(T))$, the log asset prices at $T$, conditioned on the present value $\vec{x}_0=(\log S_{1,0},\cdots,\log S_{d,0})$.
In the BS model defined by \eqref{black and scholes multi assets}, $q(\vec{x}|\vec{x}_0)$ is given by the $d$-variate normal distribution:
\begin{equation}
    q(\vec{x}|\vec{x}_0) = \frac{1}{\sqrt{(2 \pi)^d \det \Sigma}} \exp\left(-\frac{1}{2}\left(\vec{x}-\vec{\mu}\right)^T \Sigma^{-1} \left(\vec{x}-\vec{\mu}\right)\right),
\end{equation}
where $\Sigma:=(\sigma_m\sigma_n \rho_{mn}T)_{mn}$ is the covariance matrix of $\vec{x}$ and $\vec{\mu}:=\vec{x}_0+\left(rT-\frac{1}{2}\sigma_1^2T,\cdots,rT-\frac{1}{2}\sigma_d^2T\right)$.
Note that, in Eq. \eqref{expec val}, we denote the option price by $V(\vec{p})$, indicating its dependence on the parameter $\vec{p}$ such as the volatilities $\vec{\sigma}=(\sigma_1,\cdots,\sigma_d)$ and the present asset prices $\vec{S}_0$.

In FT-based option pricing, we rewrite the formula \eqref{expec val} as the integral in the Fourier space:
\begin{equation}
V(\vec{p}) = \frac{e^{-rT}}{2\pi} \int_{\mathbb{R}^d+i\vec{\alpha}}\phi(-\vec{z})\hat{v}(\vec{z})d\vec{z}. 
\label{eq:VFTBased}
\end{equation}
Here, $\vec{z}=(z_1,\cdots,z_d)$ is the wavenumber vector corresponding to $\vec{x}$.
\begin{equation}
\phi(\vec{z}):=\mathbb{E}[e^{i\vec{z}\cdot\vec{x}}|\vec{x}_0]=\int_{\mathbb{R}^d} e^{i\vec{z}\cdot\vec{x}}q(\vec{x}|\vec{x}_0)d\vec{x}    
\end{equation}
is the characteristic function, and in the BS model, it is given by
\begin{align}
    \phi(\vec{z}) = \exp \left(i \sum_{m=1}^d z_m \mu_m-
    \frac{T}{2} 
    \sum_{m=1}^d 
    \sum_{k=1}^d 
    \sigma_m \sigma_k z_m z_k \rho_{m k}\right). 
\label{eq:phi_tt}
\end{align}
$\hat{v}(\vec{z}):=\int_{\mathbb{R}^d} e^{i\vec{z}\cdot\vec{x}}v(\exp(\vec{x}))d\vec{x}$ is the Fourier transform of the payoff function $v$, and its explicit formula is known for some types of options.
For example, for a European min-call option with strike $K$, which we will consider in our numerical demonstration, the payoff function is
\begin{align}
    v_{\text{min}}(\vec{S}_T) = \max\{\min\{S_{1}(T), \ldots, S_{d}(T)\} - K, 0\}
\label{min option}
\end{align}
and its Fourier transform is \cite{Eberlein2010}
\begin{align}
    \hat{v}_{\min }(\vec{z})=-\frac{K^{1+i \sum_{m=1}^d z_m}}{(-1)^d\left(1+i \sum_{m=1}^d z_m\right) \prod_{m=1}^d (i z_m)}.
\label{eq:vmin_tt}
\end{align}
Note that for $\hat{v}_{\min }(\vec{z})$ to be well defined, $\vec{z}\in\mathbb{C}^d$ must be taken so that $\Im z_m >0$ and $\sum_{m=1}^d \Im z_m > 1$.
$\vec{\alpha}\in\mathbb{R}^d$ in Eq. \eqref{eq:VFTBased} is the parameter that characterizes the integration contour respecting the above conditions on $\Im z_m$ and taken so that $\alpha_m>0$ and $\sum_{m=1}^d \alpha_m > 1$.

In the numerical calculation of Eq. \eqref{eq:VFTBased}, we approximate it by discretization:
\begin{align}
    V(\vec{p})= \frac{e^{-rT}}{2\pi} \sum_{j_1, \cdots, j_d=-N/2}^{N/2} \phi(-\vec{z}_{\mathrm{gr},\vec{j}} - i \vec{\alpha}) \hat{v} (\vec{z}_{\mathrm{gr},\vec{j}} + i\vec{\alpha}) \Delta_{\rm vol}.
\label{eq:V_discretized}
\end{align}
Here, the even natural number $N$ is the number of the grid points in one dimension,$\vec{z}_{\mathrm{gr},\vec{j}}$ is the grid point specified by the integer vector $\vec{j}$ as
\begin{equation}
    \vec{z}_{\mathrm{gr},\vec{j}}:=(\eta j_1,\ldots,\eta j_d),
    \label{eq:zgr}
\end{equation}
where $\eta$ is the constant integration step size in each direction, and $\Delta_{\rm vol}:=\eta^d$ is the volume element. 
$\eta$ is a hyperparameter that must be appropriately determined.
In our demonstration, it is set to 0.4 or 0.3, which yields the accurate result (see Numerical demonstration for details).

\subsection{Fourier transform-based option pricing with tensor trains}

Note that to compute the sum in Eq. \eqref{eq:V_discretized}, we need to evaluate $\phi$ and $\hat{v}$ exponentially many times with respect to the asset number $d$.
This is not feasible for large $d$.
Then, to reduce the computational complexity, following Ref. \cite{kastoryano2022highlyefficienttensornetwork}, we consider approximating $\phi$ and $\hat{v}$ by TTs.
For the tensor $\phi_{j_1,\ldots,j_d}$ (resp. $\hat{v}_{j_1,\ldots,j_d}$), whose entry with index $\vec{j}$ is $\phi(-\vec{z}_{\mathrm{gr},\vec{j}} - i \vec{\alpha})$ (resp. $\hat{v} (\vec{z}_{\mathrm{gr},\vec{j}} + i\vec{\alpha})$), we construct a TT approximation $\tilde{\phi}_{j_1,\ldots,j_d}$ (resp. $\tilde{v}_{j_1,\ldots,j_d}$) by TCI.
Then, we approximately calculate Eq. \eqref{eq:V_discretized} by
\begin{equation}
    V(\vec{p}) \simeq \frac{e^{-rT}}{2\pi} \sum_{j_1, \cdots, j_d=-N/2}^{N/2} \tilde{\phi}_{j_1,\ldots,j_d}\tilde{v}_{j_1,\ldots,j_d} \Delta_{\mathrm{vol}},
\label{eq:V_TT}
\end{equation}
where each index $j_i$ ($i=1,\cdots, d$) is from $-\frac{N}{2}$ to $\frac{N}{2}$ and the corresponding integration range for each variable is $-\frac{\eta N}{2}$ to $\frac{\eta N}{2}$.
We must suitably choose both the number of grid points \(N\) and the integral step size \(\eta\), which serve as hyperparameters in this approach.

Thanks to TCI, we can obtain the approximate TTs avoiding the evaluations of $\phi$ and $\hat{v}$ at all the grid points.
Besides, given the TTs, we can compute the sum in \eqref{eq:V_TT} as the contraction of two TTs without exponentially many iterations: with the bond dimensions at most $\chi$, the number of multiplications and additions is of order $O(dN\chi^3)$.

Hereafter, we simply call this approach for FT-based option pricing aided by TTs TT-based option pricing.

\subsection{Monte Carlo-based option pricing}

Here, we also make a brief description of the Monte Carlo (MC)-based option pricing.
It is a widely used approach in practice, and we take it as a comparison target in our numerical demonstration of TT-based option pricing.

In the MC-based approach, we estimate the expectation in Eq. \eqref{expec val} by the average of the payoffs in the sample paths:
\begin{equation}
    V(\vec{p})\approx e^{-rT} \times \frac{1}{N_{\rm path}} \sum_{i=1}^{N_{\rm path}} v\left(\exp(\vec{x}_i)\right),
\end{equation}
where $\vec{x}_1,\ldots,\vec{x}_{N_{\rm path}}$ are i.i.d. samples from $q(\vec{x}|\vec{x}_0)$.
On how to sample multivariate normal variables, we leave the detail to textbooks (e.g., Ref. \cite{glasserman2004monte}) and just mention that it requires more complicated operations than simple multiplications and additions such as evaluations of some elementary functions.
Besides, calculating the payoff $v$ with the normal variable $\vec{x}_i$ involves exponentiation.
In the MC simulation for $d$ assets with $N_{\rm path}$, the number of such operations is $O(d N_{\rm path})$, and we hereafter estimate the computational complexity of MC-based option pricing by this.

\section{Learning parameter dependence with tensor trains}

\subsection{Outline}
Option prices $V(\vec{p})$ depend on input parameters $\vec{p}$ such as the volatilities $\vec{\sigma}$ and the present asset prices $\vec{S}_0$. 
In the rapidly changing financial market, these input parameters vary from time to time, which causes the change of the option price.
Therefore, if we have a function, i.e., tensor trains, that efficiently outputs an accurate approximation of the option price for various values of the input parameters, it provides a large benefit to practical business.

\begin{figure*}[ht]
    \centering
        \includegraphics[width=0.85\linewidth]{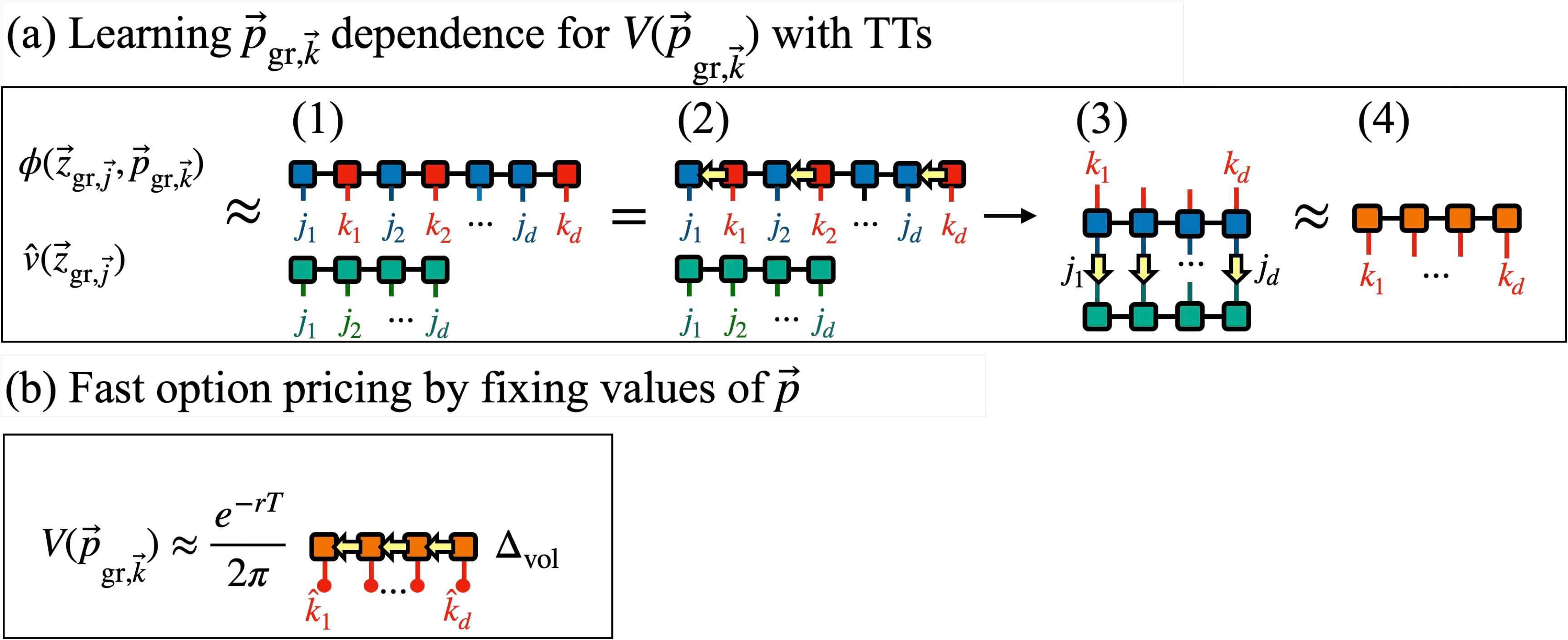}
        \caption{
        Fast option pricing based on TNs proposed in this study.
    In (a), we learn TTs from parameter-dependent function of $\phi$ and $\hat{v}$ using TCI and reduce the bond dimension of these TTs using SVD.
    We then contract the tensors associated with \(p_m\) and \(z_m\) for TT of $\phi$, resulting in TTO. 
    We then take the contraction of the TTO and TT with respect to the index vector $\vec{j}$ and compress the bond dimension of the resulting TT using SVD.
    In (b), we use this TT to perform fast option pricing for a specified parameter $\vec{p}$.}
\label{fig:new_scheme}
\end{figure*}

Then, extending the aforementioned FT-based option pricing method with TTs, we propose a new scheme to quickly compute the option price in response to the change in the input parameter set.
Using TCI, we obtain the TTs to approximate $\phi$ incorporating the parameter dependence of this function and $\hat{v}$.
The outline is illustrated in FIG.~\ref{fig:new_scheme}.
Here, focusing not all the parameters but a part of them, we take $\vec{p}$ as a $d$-dimensional vector, e.g., either of $\vec{\sigma}$ or $\vec{S}_0$.
Considering $\phi$ as the functions of not only $\vec{z}$ but also $\vec{p}$, we set in the space of $\vec{z}$ and $\vec{p}$ the grid points $(\vec{z}_{\mathrm{gr},\vec{j}},\vec{p}_{\mathrm{gr},\vec{k}})$ labeled by the index vectors $\vec{j}$ and $\vec{k}$.
Then, as illustrated in FIG.~\ref{fig:new_scheme} (a) (1), we run TCI to get the TTs $\tilde{\phi}_{j_1,k_1,\ldots,j_d,k_d}$ and $\tilde{v}_{j_1,\ldots,j_d}$ that respectively approximate the tensors $\phi_{j_1,k_1,\ldots,j_d,k_d}$ and $\hat{v}_{j_1,\ldots,j_d}$, whose entries are the values of $\phi$ and $\hat{v}$ at grid points $(\vec{z}_{\mathrm{gr},\vec{j}},\vec{p}_{\mathrm{gr},\vec{k}})$ and $(\vec{z}_{\mathrm{gr},\vec{j}})$, respectively.
We reduce the bond dimensions of these TTs using SVD. 
As shown in FIG.~\ref{fig:new_scheme} (a) (2), we then contract adjacent core tensors pairwise and obtain a tensor train operators (TTO) $\tilde{\phi}^{k_1 ,\ldots,k_d}_{j_1,\ldots,j_d}$.
Then, As shown in FIG.~\ref{fig:new_scheme} (a) (3), we contract this TTO  $\tilde{\phi}^{k_1 ,\ldots,k_d}_{j_1,\ldots,j_d}$ and TT $\tilde{v}_{j_1,\ldots,j_d}$ along the index vector $\vec{j}$ as follows:
\begin{equation}
    \tilde{V}_{k_1 ,\ldots,k_d} = 
   \sum^{N/2}_{j_1 ,\ldots,j_d=-N/2} \tilde{\phi}^{k_1 ,\ldots,k_d}_{j_1,\ldots,j_d} \tilde{v}_{j_1,\ldots,j_d},
\label{eq:TTO_contract}
\end{equation}
Then, we optimize the bond dimensions of $\tilde{V}_{k_1 ,\ldots,k_d}$ using SVD. 
Thus, as shown in FIG.~\ref{fig:new_scheme} (a) (4), we get a new function $\tilde{V}_{k_1 ,\ldots,k_d}$ whose input are parameter and output is option price.
Having this TT, we get the option price for the specified value of $\vec{p}$ as illustrated in FIG.~\ref{fig:new_scheme} (b).
By fixing the index $\vec{k}$ of $\tilde{V}_{k_1 ,\ldots,k_d}$ as $\hat{k}_1 ,\ldots,\hat{k}_d$ to the value corresponding to the specified $\vec{p}$, we get the option price for the specified $\vec{p}$ as
\begin{equation}
     V(\vec{p}_{\mathrm{gr},\vec{\hat{k}}}) \simeq
     \frac{e^{-rT}}{2\pi} 
      \tilde{V}_{\hat{k}_1 ,\ldots,\hat{k}_d} 
      \Delta_{\mathrm{vol}}.
\label{eq:option_price_TTO}
\end{equation}

Here, we mention the ordering of the local indices of the TT. 
Two core tensors in the TT that correspond to $z_j$ and $p_j$ of the same asset are arranged next to each other, i.e., the order is ($z_1 p_1 z_2 p_2 \cdots z_d p_d$). 
In the numerical demonstration, we have numerically found that this arrangement allows us to compress the TTs with parameter dependence while maintaining the accuracy of the option pricing. 
On the other hand, if the core tensors on $\vec{z}$ and those on $\vec{p}$ are completely separated, i.e.,($z_1 z_2 \cdots z_d p_1 p_2 \cdots p_d$), we have found that the accuracy of the option pricing get worse since TCI fails to learn this tensor trains. 
However, the optimal arrangement of the local indices may vary, for example, depending on the correlation matrix: intuitively, the core tensors corresponding to highly correlated assets should be placed nearby.
Although we do not discuss it in detail, this is an important topic for future research.

In the two test cases for our proposed method in Numerical demonstration, we will identify $\vec{p}$ as the volatility $\vec{\sigma}$ or the present asset price $\vec{S}_0$, the varying market parameters that particularly affect the option prices. 

\subsection{Computational complexity}
The computational complexity of TT-based option pricing, involving computation for obtaining the specific tensor components $\tilde{V}_{\hat{k}_1 ,\ldots,\hat{k}_d}$ for the fixed values of $\hat{k}_1,\ldots,\hat{k}_d$ is $O(d \chi_{\tilde{V}}^{2})$~\cite{Schollwock2011-eq}.
Here, we denote the maximum bond dimensions as $\chi_V$
In fact, the bond dimension depends on the bond index, and it is necessary to account for this for an accurate evaluation of the number of operations. 
Indeed, we consider this point in evaluating the computational complexity of TT-based option pricing demonstrated in Discussion.

Here, we ignore the computational complexity of all the processes in FIG.~\ref{fig:new_scheme} (a) and consider that in FIG.~\ref{fig:new_scheme} (b) after we get the TT only. 
This is reasonable if we can use plenty of time to perform these tasks
before we need fast option pricing.
As discussed in Discussion, we can reasonably find such a situation in practice.

\section{Numerical demonstration}\label{sec:demo}
\subsection{Details}
Now, as a demonstration, we apply the proposed method to pricing a $d$-asset European min-call option in the BS model.
In the following, we describe the parameter values used in this study, the software used, and how the errors were evaluated.

\subsubsection{Ranges of volatility and initial stock price}
With respect to $\vec{p}$, on which the TTs learn the dependence of the functions $\phi$ and $\hat{v}$, we take the two test cases: $\vec{p}$ is the volatilities $\vec{\sigma}$ or the present asset prices $\vec{S}_0$. 
In the proposed method, we need to set the range in the space of $\vec{p}$ and the grid points in it.
For each volatility $\sigma_m$, we set the range to $\sigma_m \in [0.15, 0.25)$, where the center $\sigma_m=0.2$ is a typical value of Nikkei Stock Average Volatility Index \cite{nikkei_vi} and the range width $\pm 0.05$ covers the changes in this index on most days.
For each $S_{m,0}$, we set the range to $S_{m,0} \in [90, 120)$, which corresponds to the 20\% variation of the asset price centered at 100.
The lower bound is set to not 80 but 90 because the price of the option we take as an example is negligibly small for $S_0 < 90$.
For both $\sigma_m$ and $S_{m,0}$, we set 100 equally spaced grid points in the range, and so the total number of the grid points in the space of $\vec{p}$ is $100^d$.

\subsubsection{Other parameters}
The other parameters for option pricing are fixed to the values summarized in the table below. 
The values of $\alpha$ are adopted from previous study~\cite{kastoryano2022highlyefficienttensornetwork} and we confirmed that the solution remains stable when the values are varied slightly.
$N$ and $\eta$ are set sufficiently large and small, respectively, so that the accuracy of the proposed approach becomes better than the MC.
\begin{table}[hbtp]
\centering
\scalebox{0.8}{ 
\begin{tabular}{|c|c|c|c|c|c|c|c|}
    \hline
    $T$ & $r$ & $K$ & $S_0$ & $\sigma$ & $\alpha$ & $\rho_{mn}\ (m\ne n)$ & $(N,\eta)$ \\
    \hline
    $1$ & $0.01$ & $100$ & $100$ & $0.2$ & $\frac{5}{d}$ & $\frac{1}{3}$ & 
    $\begin{cases}
        (200, 0.3) & \text{if } d = 11,\ \vec{p} = \vec{S}_0 \\
        (100, 0.4) & \text{otherwise}
    \end{cases}$ \\
    \hline 
\end{tabular}
}
\caption{
The input parameters used in the experiment, except for $\vec{\sigma}$ and $\vec{S}_0$.
When either \(\vec{\sigma}\) or \(\vec{S}_0\) is selected as the parameter dependency \(\vec{p}\), the remaining one is fixed at the value shown in the table.
For the case where $\vec{S}_0$ is varied and $d = 11$, we use $(N, \eta) = (200, 0.3)$. In all other cases, we use $(N, \eta) = (100, 0.4)$.
}
\label{tab:tt_1d_table}
\end{table}

\subsubsection{Error evaluation}
We do not have the exact price of the multivariate min-call option since there is no known analytic formula for it.
Instead, we regard the option price computed by the MC-based method with very many paths, concretely $5 \times 10^7$ paths, as the true value.
The error of the option price computed by the proposed method is evaluated by calculating the mean absolute error from the true value. As a comparison target for our method, we use an MC-based approach with \(10^6\) paths. 
As the error of the MC-based option price, we compute the half-width of its 95\% confidence interval, namely $1.96\sigma/\sqrt{n_{\rm MC}}$, where $\sigma$ is the sample standard deviation of the discounted payoff in the MC run and the path number $n_{\rm MC}$ is $10^6$.
We assess the accuracy of our method by verifying if its error is below that of the MC-based method.

Here, an issue is that the number of possible combinations of the parameter $\vec{p}$ is $100^d$, and thus we cannot test all of them.
Thus, we randomly select 100 combinations and perform option pricing for each of them. 
We compare the mean absolute error of our method for the 100 parameter sets with the one obtained from the Monte Carlo simulations with the same parameter setting. 

\subsubsection{Software and hardware used in this study}
\texttt{TensorCrossInterpolation.jl}~\cite{xfac} was used for learning tensor trains from functions appearing Fourier-based option pricing.
The Monte Carlo simulations were carried out using \texttt{tf-quant-finance}~\cite{monte}. Parallelization was not employed in either case. 
GPUs were not utilized in any of the calculations. 
The computations were performed on a 2023 MacBook Pro featuring a 12-core Apple M2 Max processor and 32GB of 400GB/s unified memory.

\subsection{Results}

We show the results for the computational complexity, time, and accuracy of TT-based and MC-based option pricing when two parameters $\vec{\sigma}$ and $\vec{S_0}$ are varied.

\begin{table*}[htbp]
  \centering
  \begin{subtable}[h]{0.74\textwidth}
    \centering
    \caption{$\vec{\sigma}$}
    \scalebox{1.0}{
      \begin{tabular}{|c|c|c|c|c|c|c|c|c|c|}
        \hline
        $d$ & $e_{\mathrm{TT}}$ & $e_{\mathrm{MC}}$ & $c_{\mathrm{TT}}$ & $c_{\mathrm{MC}}$ & $t_{\mathrm{TT}}\,[$s$]$ & $t_{\mathrm{MC}}\,[$s$]$ & $\chi_{\phi}$ & $\chi_{\hat v}$ & $\chi_{\tilde V}$ \\
        \hline
         5 & 0.00178 & 0.00606 & 16 & $5.0\times10^{6}$ & $4.91\times10^{-7}$ & $2.84\times10^{-1}$ & 16 & 11 & 2 \\
         6 & 0.00154 & 0.00503 & 20 & $6.0\times10^{6}$ & $6.37\times10^{-7}$ & $3.38\times10^{-1}$ & 16 & 11 & 2 \\
         7 & 0.00134 & 0.00428 & 24 & $7.0\times10^{6}$ & $9.12\times10^{-7}$ & $3.71\times10^{-1}$ & 17 & 11 & 2 \\
         8 & 0.00136 & 0.00372 & 28 & $8.0\times10^{6}$ & $1.75\times10^{-6}$ & $4.09\times10^{-1}$ & 18 & 11 & 2 \\
         9 & 0.000867 & 0.00329 & 32 & $9.0\times10^{6}$ & $2.47\times10^{-6}$ & $4.51\times10^{-1}$ & 20 & 11 & 2 \\
        10 & 0.00229 & 0.00294 & 10 & $1.0\times10^{7}$ & $1.18\times10^{-6}$ & $5.11\times10^{-1}$ & 20 & 11 & 1 \\
        11 & 0.000554 & 0.00265 & 11 & $1.1\times10^{7}$ & $1.16\times10^{-6}$ & $5.04\times10^{-1}$ & 20 & 11 & 1 \\
        \hline
      \end{tabular}
    }
  \end{subtable}
  \\
  \vspace{10pt}
  \begin{subtable}[h]{0.74\textwidth}
    \centering
    \caption{$\vec{S}_0$}
    \scalebox{1.0}{
      \begin{tabular}{|c|c|c|c|c|c|c|c|c|c|}
        \hline
        $d$ & $e_{\mathrm{TT}}$ & $e_{\mathrm{MC}}$ & $c_{\mathrm{TT}}$ & $c_{\mathrm{MC}}$ & $t_{\mathrm{TT}}\,[$s$]$ & $t_{\mathrm{MC}}\,[$s$]$ & $\chi_{\phi}$ & $\chi_{\hat v}$ & $\chi_{\tilde V}$ \\
        \hline
         5 & 0.00151 & 0.00773 & 16 & $5.0\times10^{6}$ & $7.67\times10^{-7}$ & $3.99\times10^{-1}$ & 18 & 11 & 2 \\
         6 & 0.00122 & 0.00640 & 20 & $6.0\times10^{6}$ & $6.34\times10^{-7}$ & $5.15\times10^{-1}$ & 19 & 11 & 2 \\
         7 & 0.00112 & 0.00547 & 24 & $7.0\times10^{6}$ & $9.21\times10^{-7}$ & $5.37\times10^{-1}$ & 21 & 10 & 2 \\
         8 & 0.000973 & 0.00477 & 28 & $8.0\times10^{6}$ & $1.63\times10^{-6}$ & $6.18\times10^{-1}$ & 23 & 11 & 2 \\
         9 & 0.000686 & 0.00424 & 32 & $9.0\times10^{6}$ & $1.19\times10^{-6}$ & $7.95\times10^{-1}$ & 24 & 11 & 2 \\
        10 & 0.000662 & 0.00377 & 36 & $1.0\times10^{7}$ & $1.47\times10^{-6}$ & $8.77\times10^{-1}$ & 24 & 11 & 2 \\
        11 & 0.00114  & 0.00339 & 40 & $1.1\times10^{7}$ & $1.93\times10^{-6}$ & $8.43\times10^{-1}$ & 25 & 13 & 2 \\
        \hline
      \end{tabular}
    }
  \end{subtable}
  \caption{
    The results of TT-based option pricing incorporating (a) $\vec{\sigma}$ and (b) $\vec{S}_0$ dependence.
    Here, we set the ranges $\sigma_m \in [0.15,0.25)$ and $S_{m,0}\in[90,120)$, placing 100 equally spaced grid points within each range.
    $t_{\mathrm{TT}}$ and $t_{\mathrm{MC}}$ are averages over 100 measurements.
  }
  \label{table:table_res}
\end{table*}

The results are summarized in TABLE~\ref{table:table_res}. 
In particular, the computational complexity versus $d$ is plotted in FIG.~\ref{fig:table_cc}.
The mean absolute error in TT-based option prices among runs for 100 random parameter sets is represented by $e_{\mathrm{TT}}$. 
For the same parameter sets, we compute mean values of the absolute values of the error in the MC-based option price with $10^6$ paths and denote it by $e_{\mathrm{MC}}$.
The computational complexity and time of TT-based option pricing are represented by $c_{\mathrm{TT}}$ and $t_{\mathrm{TT}}$, respectively, and those of MC-based pricing with $10^6$ paths are denoted by $c_{\mathrm{MC}}$ and $t_{\mathrm{MC}}$.
To maintain the desired accuracy of option pricing, we set the tolerance of TCI sufficiently low, concretely $\epsilon_{\mathrm{TCI}} = 10^{-9}$.
Subsequently, we reduce the bond dimension by SVD, with
the tolerance of SVD set to  $\epsilon^{\phi,\hat{v}}_{\mathrm{SVD}} = 1.0 \times 10^{-6}$ for $\phi$ and $\hat{v}$.
The maximum bond dimensions of the TTs for $\phi$ and $\hat{v}$ are denoted as $\chi_{\phi}$ and $\chi_{\hat{v}}$, respectively, in TABLE~\ref{table:table_res}.
In the SVD on \(\tilde{V}_{k_1, \ldots, k_d}\) obtained by contracting the TTs for $\phi$ and $\hat{v}$, we set the tolerance $\epsilon^{\tilde{V}}_{\mathrm{SVD}}= 1.0 \times 10^{-6}$, and the maximum bond dimension of the resulting TT is denoted by \(\chi_V\).

\subsubsection{The case of varying volatilities}
\begin{figure*}[htbp]
    \centering
\includegraphics[width=1.0\linewidth]{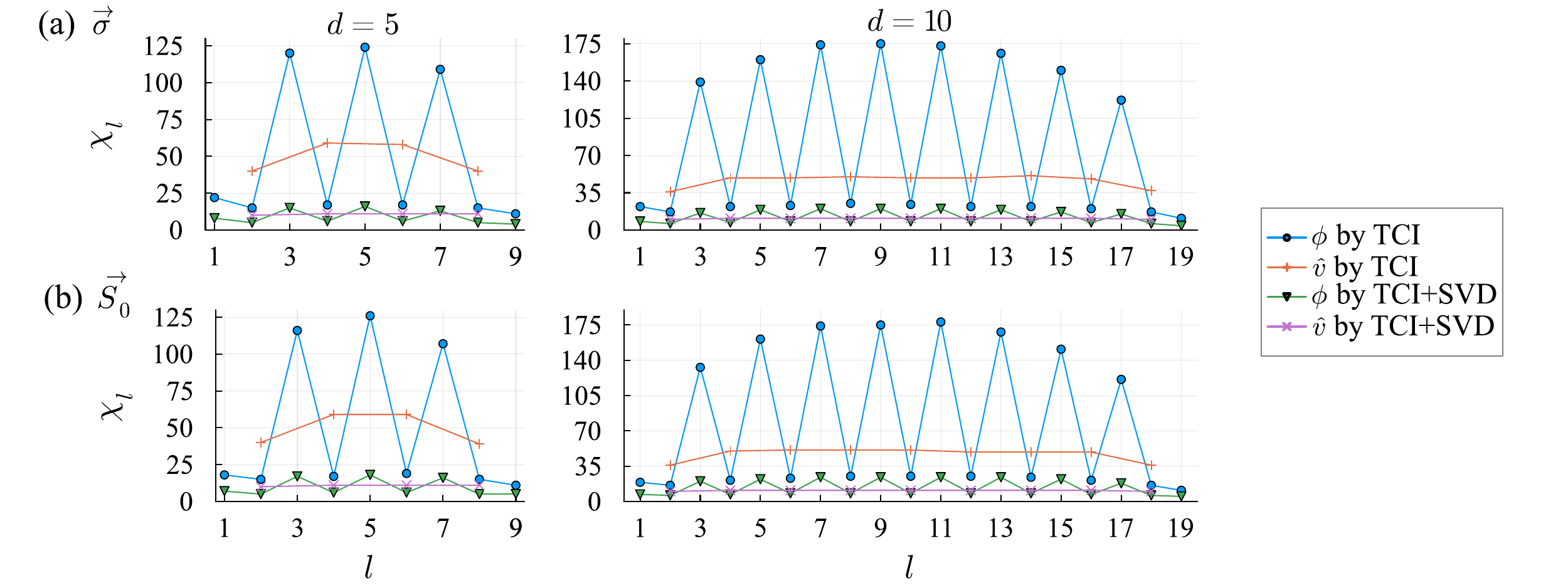}
        \caption{
        The bond dimensions $\chi_{l}$ of each bond $l$ for $\phi$ and $\hat{v}$, obtained through TCI and SVD, incorporating dependencies on (a) $\vec{\sigma}$ and (b) $\vec{S}_0$. 
        The odd bond $l=2i-1$ connects the core tensors for $j_m$ and $k_m$, the indexes for $z_m$ and $p_m$, respectively, and the even bond $l=2i$ connects those for $k_m$ and $j_{m+1}$.
        Note that $\hat{v}$ does not depend on parameters, and thus the bond $l$ takes only even values $2i$ (connecting the core tensors for $z_m$ and $z_{m+1}$).
        It is noteworthy that the graph of bond dimensions exhibits a characteristic jagged shape. 
        The bond dimensions between sites for $j_m$ and $k_{m}$, which are related to the same asset, are large, and the ones between sites for $k_m$ and $j_{m+1}$, which are related to different assets, are small. 
       }
\label{fig:bonddim}
\end{figure*}

TABLE~\ref{table:table_res} (a) shows the computational results of TT-based option pricing when we consider $\vec{\sigma}$ dependence.
TT-based option pricing demonstrates advantages in terms of computational complexity and time over the MC-based method.
The bond dimensions of the TT results for $d=5$ and $d=10$ are depicted in FIG.~\ref{fig:bonddim}(a). 
By applying SVD to the TTs trained via TCI, the bond dimensions between tensors related to $p_m$ and $z_{m+1}$ could generally be maintained at around 10.
The details of this compression by SVD are described in \ref{appendix:svd_tol}.
The maximum bond dimension $\chi_{\tilde{V}}$ of $\tilde{V}_{k_1 ,\ldots,k_d}$ is maintained at 2, or 1  especially when $d=10,11$. 
Therefore, the computational complexity is much lower than that of MC-based option pricing.
Besides, the error in the TT-based result is smaller than that in the MC-based result in all the tested cases.

\subsubsection{The case of varying initial stock prices}
TABLE~\ref{table:table_res} (b) shows the results of TT-based option pricing when we consider $\vec{S_0}$ dependence.
TT-based option pricing demonstrates superiority over the Monte Carlo method in terms of computational complexity and time. 
For $d$=5 and 10, after compression using SVD and contractions, the bond dimensions between tensors related to $p_m$ and $z_{m+1}$ were reduced to around 10 (refer to FIG.~\ref{fig:bonddim}(b)).
The maximum bond dimension, \(\chi_{\tilde{V}}\), of \(\tilde{V}_{k_1, \ldots, k_d}\) is 2.
Therefore, we saw superiority over the Monte Carlo method in terms of computation complexity and time.

Compared with the MC-based method with $10^6$ paths, the TT-based mathod again gives the smaller error in every value of $d$.

\subsubsection{Randomness in learning the TTs}
Here, we mention the randomness in learning the TTs and the error induced by it.
We should note that TCI is a heuristic method, and depending on the choice of the initial points, the learning might not work well. 
That is, the error defined by Eq. \eqref{eq:errTCI} might not go below the threshold. 
To assess such a fluctuation of the accuracy, for \(d=10\) in the case of varying $\vec{\sigma}$, we evaluated the mean and standard deviation of the accuracy of the TT-based method in 20 runs, with initial points randomly selected in each. 
As a result, the mean was 0.00230 and the standard deviation was $4.74 \times 10^{-9}$, which indicates that the accuracy fluctuation of TT-based option pricing is very small.

\subsubsection{Total computational time for obtaining the TTs}
We mention the total computational time for obtaining the TTs for $\phi$ and $\hat{v}$ and $\tilde{V}_{k_1 ,\ldots,k_d}$ through TCI and SVD.
We focus on TCI because it dominates over SVD.
It took about $12.5$ minutes for $d=11$ in the case that $\vec{S}_{0}$ dependence is involved.
This is sufficiently short for a practical use-case of the TT-based method mentioned exemplified in Discussion.
We also note that, when we do not incorporate the parameter dependence into the tensor train for $\phi$ as in Ref.~\cite{kastoryano2022highlyefficienttensornetwork}, TCI takes a much longer time than the MC-based method, 7.3 seconds for $d=11$ in the case of $\vec{\sigma}$ dependence. 
In this setup, we set $\eta=0.48$, $\epsilon_{\mathrm{TCI}}=10^{-6}$, $N=50$ and all other parameters are the same as those in the default settings listed in Table~\ref{tab:tt_1d_table}.
This means that running TCI every time the parameter varies does not lead to the time advantage of the TT-based method over the MC-based one.


\section{Discussion}\label{sec:summary}
We propose a method that employs a single TCI to learn TTs incorporating parameter dependence from the function of Eq.~\eqref{eq:phi_tt}, enabling fast option pricing in response to varying parameters. 
In this study, we considered scenarios with varying volatility and present stock prices as benchmarks for our proposed method. 
Up to \(d=11\), we demonstrated superiority in both computational complexity and time.
Note that in the MC-based method, the implementation may have room for improvement to reduce computational complexity and time.
We have also seen that in all the tested cases, the error in the TT-based result is smaller than that of the MC-based result with \(10^6\) paths. 

Now, let us consider how the proposed method provides benefits in practical business in financial firms.
An expected way to utilize this method is as follows.
At night, when the financial market is closed, we learn the TTs and perform contractions, and then, during the day, when the market is open, we use the TT of parameter-dependent option prices to quickly price the option for the fluctuating parameters.
If we pursue the computational speed in the daytime and allow the overnight precomputation to some extent, the above operation can be beneficial.
In light of this, it is reasonable that we compare the computational complexity in the TT-based method after the TTs are obtained with that in the MC-based method, neglecting the learning process and contraction.

Finally, we discuss future research directions. 
For our method to be more practical, compressing many kinds of input parameters including both $\vec{\sigma}$ and $\vec{S}_0$ into a single TT is desired.
For example, taking into account the dependence on the parameters concerning the option contract, such as the maturity $T$ and the strike $K$, enables us to price different option contracts with a single set of TTs.
Although this is a promising approach, there might be some issues. For example, it is non-trivial whether TTs incorporating many parameter dependencies have a low-rank structure. Thus, we will leave such a study for future work. 
Besides, to enhance the practical benefit, we should extend our methodology so that it is applicable to more advanced settings in option pricing.
With respect to the pricing model, while this paper has considered the BS model, more advanced models such as the local volatility model~\cite{dupire1994pricing}, stochastic volatility model~\cite{heston1993closed,hagan2002managing} and Lévy models~\cite{RePEc:ucp:jnlbus:v:63:y:1990:i:4:p:511-24, RePEc:spr:finsto:v:2:y:1997:i:1:p:41-68} are also used in practice.
Applying our method to such models, where the dependencies on the parameters in the models are incorporated into TTs, is an interesting direction.
Expanding the scope to the broad types of products is also desired.
While this paper focused on European-type options, more complicated ones such as American and Bermudan options are also traded widely, and pricing of such options is more time-consuming.
Considering whether our method can be extended for such options will be an important and interesting challenge.

\section{Acknowledgements}
We are grateful to Marc K. Ritter, Hiroshi Shinaoka, and Jan von Delft for providing access to \texttt{TensorCrossInterpolation.jl} for TCI~\cite{xfac}.
We deeply appreciate Marc K. Ritter and Hiroshi Shinaoka for their critical remarks on the further improvement of our proposed method.
R. S. is supported by the JSPS KAKENHI Grant No. 23KJ0295. 
R. S. thanks the Quantum Software Research Hub at Osaka University for the opportunity to participate in the study group on applications of tensor trains to option pricing.
R.S. is grateful to Yusuke Himeoka, Yuta Mizuno, Wataru Mizukami, and Hiroshi Shinaoka since R.S. got the inspiration for this study through their collaboration.
K. M. is supported by MEXT Quantum Leap Flagship Program (MEXT Q-LEAP) Grant no. JPMXS0120319794, JSPS KAKENHI Grant no. JP22K11924, and JST COI-NEXT Program Grant No. JPMJPF2014.
This work was supported by the Center of Innovation for Sustainable Quantum AI (JST Grant Number JPMJPF2221).

\appendix
\section{How to set the tolerance $\epsilon^{\phi,\hat{v}}_{\mathrm{SVD}}$}\label{appendix:svd_tol}
\begin{figure*}[]
    \centering
        \includegraphics[width=0.5\linewidth]{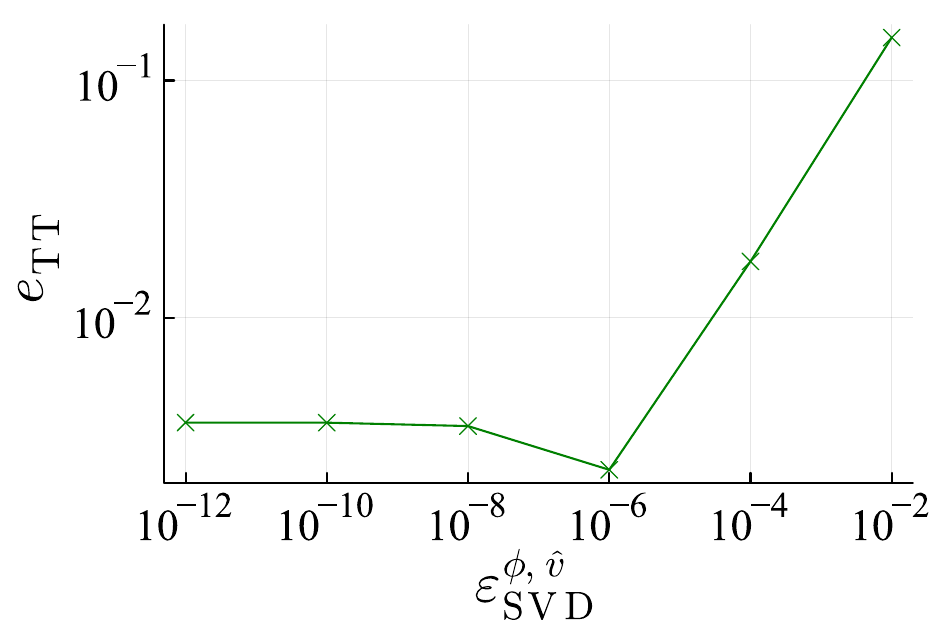}
        \caption{The maximum error $e_{\mathrm{TT}}$ when applying SVD with various $\epsilon^{\phi,\hat{v}}_{\mathrm{SVD}}$ values to the TTs under the parameter settings of $d = 10, \vec{\sigma}, \eta = \text{auto}$. 
        }
\label{fig:each_sigma_100_svd_relerr_avedim}
\end{figure*}
Figure \ref{fig:each_sigma_100_svd_relerr_avedim} shows the maximum relative error $e_{\mathrm{TT}}$ when applying SVD with varying $\epsilon^{}_{\mathrm{SVD}}$ to the TT with parameters for $d = 10,~\vec{\sigma}$, obtained through TCI. 
Here, we fixed the tolerance of SVD $\epsilon^{\tilde{V}}_{\mathrm{SVD}}$ to $10^{-9}$.
The tolerance $\epsilon^{\phi,\hat{v}}_{\mathrm{SVD}}$ for SVD was chosen to maintain $e_{\mathrm{TT}}$ smaller than $e_{\mathrm{MC},10^5}$.
From FIG.~\ref{fig:each_sigma_100_svd_relerr_avedim}, it can be seen that $e_{\mathrm{TT}}$ increases sharply between $\epsilon_{\mathrm{SVD}} = 10^{-6}$ and $10^{-5}$, suggesting that setting $\epsilon_{\text{SVD}}$ around $5.0 \times 10^{-6}$ is appropriate for keeping $e_{\mathrm{TT}}$ smaller than $e_{\mathrm{MC},10^5}$.
By compressing the bond dimension with SVD, the computational time for the contraction of these optimized TTs also can be kept low.

From the fact that we can keep the maximum relative error $e_{\mathrm{TT}}$ small through SVD, it is suggested that the TTs obtained by TCI contain redundant information. By using SVD to get an optimal approximation in terms of the Frobenius norm, the reduced redundant information could be effectively removed. 
In addition, it is surprising that in the analysis with $\epsilon_{\text{SVD}} = 10^{-6}$, the maximum relative error decreased compared to before compression by SVD.
We consider that the error contained in TTs obtained by TCI is eliminated through SVD by chance.
We expect that this phenomenon does not occur generally, and in fact, it did not occur for other asset numbers or parameters.

\bibliographystyle{apsrev4-1}
\bibliography{ref}

\end{document}